\documentstyle[aps,prc,prabib]{revtex}
\begin{document}
\def\s{$\,$}
\title{Nuclear Shell Structure and
Chaotic Dynamics in Hexadecapole Deformation}
\author{$^{\dagger}$W.D. Heiss, $^{\dagger}$$^{\ddagger}$R.G. Nazmitdinov
and $^{\dagger}$S. Radu}
\address{$^{\dagger}$Centre for Nonlinear Studies and Department of Physics\\
University of the Witwatersrand,
P.O. Wits 2050, Johannesburg, South Africa\\
$^{\ddagger}$Bogoliubov Laboratory of Theoretical Physics, JINR,
Dubna 141980, Russia}
\maketitle
\begin{abstract}
The effect of an axially symmetric hexadecapole term is investigated in a
strongly deformed quadrupole potential. While the system is nonintegrable
and shows significant chaotic behaviour classically, the quantum mechanical
treatment not only produces a general smoothing effect with regard to chaos
but even yields a pronounced shell structure at certain hexadecapole
strength parameter values for oblate and prolate deformation.
\end{abstract}
\par
\vspace{0.2in}
PACS numbers: 21.60 Cs, 05.45.+b \par
\vspace{0.2in}
 Recently, the occurence of shell structure has been reported for
 many body
 systems like nuclei  and  metallic clusters \cite{Ar,HNR} at strong
octupole deformation. A major conclusion of \cite{HNR} is
that, albeit nonintegrable,
an octupole admixture to quadrupole oscillator potentials
leads, for some values of the octupole strength,
to a shell structure similar to a plain but more deformed quadrupole
potential. This feature shows that there is a tendency of the system to
restore the full original symmetry which is destroyed by the octupole
term. In other words, we encounter the restoration of the original symmetry
under certain conditions despite the system being nonintegrable.

In an integrable case the higher symmetry can cause a high degree of
degeneracy of the eigenstates. Consequently, a spherical symmetry
leads to very strong shell effects manifested in the  stability of the noble
gases, metallic clusters and magic nuclei. However, non-spherical shapes
may be preferred by the system: shell effects associated with lesser
symmetry occur. In super- and hyperdeformed nuclei deviations from
the spherical shape are a consequence
of strong shell closures giving rise to largest level bunching (largest
degeneracy or lowest level density) for particle numbers where the
spherical shell would be only partially filled.
The concept of pseudo-SU(3) symmetry \cite{Ra,Dra} has been introduced
\cite{Dud7} to explain many features of superdeformed states described
within a realistic nuclear potential. In the new
scheme the spin-orbit splitting appears very small and the properties of
the single-particle spectrum are similar to those observed in the
three-dimensional harmonic oscillator with rational ratios of the frequencies
(RHO) \cite{Rag,NaD}. Therefore, one can argue that the
single-particle shell structure of RHO (which is an integrable system) should
reflect the essential properties of super- and hyperdeformed nuclei.

 The need for multipole deformations higher than the quadrupole has been
recognized in nuclei and metallic clusters in numerous calculations to
explain experimental data. The most important ones are the octupole and
hexadecapole terms. Inclusion of either term leads to a nonintegrable problem.
The hexadecapole deformation is essential for the understanding
of equilibrium shapes and the
fission process of super- and hyperdeformed systems \cite{Ab,Dud}.
In the case of metallic clusters, the axial
hexadecapole deformation is important for the interpretation of experimental
data in simple metals \cite{HB}.
Usually the even and odd multipoles are considered together.
However, obvious differences between the octupole and hexadecapole term
suggests a separate study of their respective effects
to shell structure phenomena.
We aim at shedding more light on this old question from a new point of view.
Our approach is based on the
connection between shell structure phenomena in the quantum
spectrum and ordered motion in the classical analoguous case.
Shell structures in the quantum mechanical spectrum are associated
with periodic orbits in the corresponding classical problem
\cite{BB,BM,SM,HN}. The periodic orbits are associated with invariant tori
of the Poincar\'e sections. If the classical
problem is chaotic, the invariant tori disintegrate or disappear, and
the shell structure of the quantum spectrum is affected by the degree of
chaos \cite{US2,AM}. In this paper
particular emphasis is placed upon strongly quadrupole deformed systems like
superdeformed nuclei, and we concentrate
only on the most important even multipoles, quadrupole and hexadecapole
deformations. In a previous analysis of an octupole term \cite{HNR}
it was demonstrated that shell structure can exist only
for the super- and more deformed prolate case. In this paper we resume the
study of quantum-classical correspondence for the hexadecapole term and
show the occurrence of shell structure in prolate as well as oblate systems.

We investigate the classical and quantum
mechanical single-particle motion in an axially symmetric harmonic
oscillator potential including the hexadecapole term, {\it viz.}
\begin{equation}
V(\varrho,z) = \frac{m\omega^2}{2} \left(
\varrho^2 + \frac{z^2}{b^2} + \lambda
\frac{8z^4 - 24 z^2\varrho^2 + 3\varrho^2}{z^2 + \varrho^2} \right)
\end{equation}
where cylindrical coordinates $(\varrho,z)$ are used.
The quadrupole deformation and the hexadecapole
strength are denoted by $b$ and $\lambda$, respectively.
We use for the hexadecapole term the expression $r^2P_4(\cos \theta)$
to ensure a proper bound-state problem
provided the hexadecapole strength is restricted to the range:
\begin{equation}
{\rm max}\left( -\frac{1}{8b^2},-\frac{1}{3} \right) <
\lambda < \frac{(4b^2+3)\sqrt{5} + 2\sqrt{22b^4+24b^2+7}}{48\sqrt{5}} .
\end{equation}
Since the potential scales as $V(\beta\varrho,\beta z)=
\beta^2 V(\varrho,z)$, only one energy value has
to be considered \cite{LL}.
The axial symmetry guarantees conservation of the $z$-component of the
angular momentum $p_\varphi$, and the $z$-projection $m$ of the
angular momentum is therefore a good quantum number.

The classical perturbative treatment follows the secular perturbation
theory \cite{Lib} using the {\it removal of resonances} method (RRM);
the application to quadrupole deformed systems including an octupole term
is described in detail in \cite{HNR}. Notice, that the RRM is particularly
effective in the case of a two-dimensional problem. The main task is to find
the values of the parameters $b$ and $\lambda$ for which the nonintegrable
problem reduces effectively to an integrable case, which is in our case
the two-dimensional oscillator with new effective frequencies.
In this way we obtain good estimates for the winding numbers of the classical
orbits and find the conditions where the original symmetry is restored.
In contrast to the octupole case the hexadecapole deformation evokes more
chaotic behaviour and involves resonances of higher order, hence the method
may not  produce results of comparable accuracy.
However, without major efforts the approach turns out to give
reliable guide lines as to what to expect quantum mechanically.

In the oblate case, the perturbative method yields an unperturbed motion in
the $z$-coordinate with a frequency $\omega_z =\omega/b$.
The motion in $\varrho$ is described by the effective potential:
\begin{equation}
 V^{\rm obl}_{\rm eff}(\varrho, \xi_z) =\frac{m\omega^2}{2} \left(
 (1-32\lambda)\varrho^2 +
 35\lambda \frac{\varrho^3{\rm sign}(\varrho)}{\sqrt{\varrho^2+\xi_z^2}}
 + 4\lambda\xi_z^2\right)+{p_\varphi ^2\over 2m\varrho ^2}
 \label{eq:1}
\end{equation}
where $\xi_z= \sqrt{2bE_z/(m\omega^2)}$ reminds us of the actually coupled
motion in $\varrho$ and $z$; it represents the portion $E_z$ residing in
the $z$-motion of the total energy. From Eq.(\ref{eq:1}) the frequencies
of the anharmonic oscillations $\omega_\varrho$ can be evaluated.
It turns out that, if the quadrupole deformation is sufficiently large,
for instance $b=2/5$, the winding number defined as $\omega_\varrho/\omega_z$
is independent of $\xi_z$ for about $65\%$ of its allowed range. In fact,
for $0\le \xi_z \le 0.65\xi_z^{{\rm max}} $ and $p_\varphi =0$ it is close to
$\omega_\varrho/\omega_z = b\sqrt{1+3\lambda}$ which is the exact expression
for $\xi_z =0$. For $b=2/5$ and
$\lambda = 0.18$, we obtain $\omega_\varrho/\omega_z = 1:2$ which applies for
about two thirds of the $\xi_z$ range. For these values of the deformation
parameters we therefore expect in the quantum spectrum a sequence of
levels which have virtually the same pattern as the shell structure of a
pure oblate superdeformed $(b=1/2)$ system. A slight $p_\varphi $-dependence
of the winding number decreases the actual value of $\lambda $ (see below).

A Strutinsky-type analysis has been carried out for levels
comprising values of $m$ from 0 to 15. The total energy
$E_{\rm tot}(\lambda,N)$ which is the sum over all single particle levels
up to $N$  is fitted by a polynomial
$E_{\rm smooth}(\lambda,N)= \sum_{i=0}^{4}c_i(\lambda)N^{i/3}$, and the
fluctuation $\delta E(\lambda,N) = E_{\rm tot}(\lambda,N) -
E_{\rm smooth}(\lambda,N)$ is plotted versus $\lambda$
and $N$. The resulting contour plot is displayed in Fig.(1a).

We discern significant alternating minima and maxima along
the line $\lambda\approx 0.09$ clearly indicating shell structure.
The $p_\varphi $-dependence of the winding number is reflected in a slight
$m$-dependence of the spectrum. This is, however, sufficiently
weak so as not to disturb the bunching of the levels when all $m$-values
are considered. The $m$-dependence is noted in a decrease of the effective
$\lambda $-value where shell structure occurs and in a slight
broadening of the peaks in Fig.(2); we find bunching of levels for $m=0$ at
$\lambda \approx 0.12$ and for $m=15$ at $\lambda \approx 0.07$.

Apart from the shell structure there are some remarkable stability
islands for $\lambda\approx$ 0.25, 0.22, 0.21 corresponding to the particle
number $N$ (or $Z)\approx$ 56, 86 and 126, respectively. Their physical
significance should be subjected to further experimental investigation.
{}From the quantity $\Delta E(\lambda ,N)=\delta E(\lambda,N+1) +
\delta E(\lambda,N-1)-2\delta E(\lambda,N)$ which is displayed in Fig.(2a)
we obtain the precise location of the magic numbers. The pronounced peaks
coincide with the magic numbers of the oblate superdeformed ($b=1/2$) system.

While the quantum spectrum shows a fair degree
of order, the analogous classical problem reveals a significant amount of
chaos. In Fig.(3) we display surfaces of section in the
$(\varrho,p_\varrho)$-plane for $p_\varphi=0$; $p_\varphi>0$ does not lead
to further insight. The phase space accommodates
the coexistence of chaotic and regular motion. It is dominated by a large
fourfold separatrix which contains in its four outer centres the four
stability islands of the stable periodic orbit with winding number 1:2. It is
this orbit which is responsible for the shell structure in the quantum
spectrum. Note that the separatrix occupies about two thirds of phase space.
The decay of the separatrix is clearly discernible. In the centre, additional
short periodic orbits with winding numbers $2:5, 1:3, 2:1$ are found. They
can be associated with minor peaks in the power spectrum of the levels
at $\lambda=0.12$  shown in Fig.(4). Yet, their contribution to the level
density of the low lying levels $(N\leq 150)$ is less pronounced than the
contribution of the main orbit with its winding number $1:2$. Note the
oscillatory envelope of the major peak which is in line with general
expectations \cite{hemu}. In summary, we
find that neither the chaotic orbits nor the abundance of periodic orbits
are significantly reflected in the quantum spectrum except for the shortest
one dominating the phase space. The model constitutes a fine example of
quantum suppression of classical chaos.

The same analysis can be carried out for prolate systems.
The estimated winding number $\omega_\varrho/\omega_z$, which is now a
function of $\xi_\varrho=\sqrt{2E_{\varrho }/(m\omega^2)}$, is equal to its
value at $\xi_\varrho=0$ for about
85\% of the allowed range for $\xi_\varrho$, provided the quadrupole
deformation $b$ is sufficiently large. At $\xi_\varrho=0$, the winding number
becomes $\omega_\varrho/\omega_z = b/\sqrt{1+8b^2\lambda}$. For $b=5/2$ and
$\lambda =0.011$ this yields the winding number 2:1.
The Strutinsky-type analysis displayed in Figs.(1b) and (2b) nicely confirms
the quantum mechanical expectation, in fact, this time the RRM
gives a very accurate prediction of $\lambda$ for the occurence of
superdeformed shell structure. There is no noticeable angular momentum
dependence in the prolate case, since the $\varrho$-motion is
virtually unperturbed.
The hexadecapole strength considered is much smaller  than in the
oblate case and the shell structure occurs
over a relatively wider range of $\lambda$ values with magic numbers for $N$
(or $Z)\approx$ 40, 80, 140.
This result agrees with a prediction of superdeformed shell structure for
prolate nuclei \cite{RS} which confirms the physical relevance of our
analysis. The numerical integration of the equations of motion reveals the
same kind of structure as in the oblate case.
The major orbit with the winding number $2:1$ lies in the outer centres of
the four-fold  separatrix that occupies a large portion of the phase space.
Other short periodic orbits which are insignificant for the quantum-classical
correspondence ($N\leq 150$) again occur in the surface of sections.
This time the onset of chaos along the four-fold separatrix is less
pronounced.

One of the major results of this work is the occurrence of shell structure
in an oblate deformed potential when a hexadecapole term is added. This is
not an obvious result in view of the nonintegrability of the problem. Also,
the occurrence of shell structure in the oblate case is of particular
interest as the adding of an octupole term produces chaos without structure.
The explanation for the latter result is found in the RRM which excludes,
within the model considered, the possibility of decoupling the two degrees of
freedom for odd multipoles in the oblate case; only the addition of even
multipoles can restore the original symmetry and give rise to shell structure.

The Strutinsky-type analysis indicates stable energy configurations
for oblate and prolate deformation; it appears that for particular
magic numbers the oblate configuration is favoured as the respective minima
in Fig.(1) are more pronounced. For positive
values of $\lambda $ the net effect for the spectrum lies in its resemblance
to that of a lesser quadrupole deformation. We mention that a negative value
of $\lambda $ leads to the opposite effect in that a spectrum similar to
that of a hyperdeformed system is found.
Note that for prolate deformation
the hexadecapole term is much weaker than that in the oblate case, i.e.
a $\lambda $-value as large as the one used in the oblate case would produce
hard chaos in the prolate case with all shell structure destroyed.
In fact, a minute admixture of a hexadecapole deformation in the prolate
case changes the pure $b=5/2$ deformation into a $b=2$ situation for the lower
part of the energy spectrum relevant for nuclei. This means, that shell
structure manifested experimentally by specific magic numbers
cannot be directly associated with a definite type of deformation
of the system. For instance, within our model similar superdeformed shell
patterns can be reproduced in the prolate case with a combination of
less quadrupole + octupole deformations \cite{HNR} or with more deformed
quadrupole system and a hexadecapole admixture.
Experimental information extracted from electromagnetic transitions
could clarify as to whether we are faced with the plain superdeformed
or the stronger deformed system which includes a hexadecapole admixture,
or the less deformed system but with octupole deformation.
The question as to whether or not the combination of an octupole and
a hexadecapole term destroys shell structure in the oblate and enhances
shell structure in the prolate case, is under investigation.

Finally we
stress that even though the quantum mechanical treatment shows a certain
degree of suppression of classical chaos, the occurrence of a new shell
structure which differs from the unperturbed case is clearly brought
about by the nonlinear character of the problem; the pattern emerges when
the original symmetry of the unperturbed Hamiltonian is restored even though
the problem becomes nonintgrable due to the additional term.

R.G.N. acknowledges financial support from the Foundation for Research
Development of South Africa.

\newpage
\centerline{\bf Figure Captions}

{\bf Fig.1}  Contour plots of the fluctuating part of the total energy
$\delta E$ as a function of particle number $N$ and hexadecapole
strength $\lambda$ for (a) the oblate ($b=2/5$) and (b) the prolate ($b=5/2$)
case. Dark regions correspond to energy minima which are associated to shell
closures. Volume conservation is taken into account.

{\bf Fig.2}  The quantity $\Delta E(\lambda,N)$ (see in the text)
 for (a) the oblate ($b=2/5,\lambda=0.09$) and (b) the prolate case
 ($b=5/2,\lambda=0.011$). The highest peaks occur at magic
 numbers which coincide with those of
 the oblate superdeformed and the prolate superdeformed systems,
 respectively.

{\bf Fig.3}  Surface of sections indicating the phase space structure in
the oblate situation for $p_{\varphi }=0$ and $\lambda =0.12$.

{\bf Fig.4} Power spectrum of the level density for $\lambda=0.12$,
$m=0$ and $b=2/5$. The period of the peaks marked by a diamond agrees
perfectly with that of the classical 1:2 periodic orbit. It is the
orbit associated with the magic numbers of Fig.(2a) and its shape is
displayed in the insert.

\end{document}